# Using Arduino in Physics Experiments: Determining the Speed of Sound in Air

**Atakan Çoban[1] and Niyazi Çoban[2]**

[1]Physics Department, Yeditepe University, Turkey.
[2]Serdarlı Primary School, Serdarlı, Cyprus.

E-mail: atakancoban39@gmail.com

**Abstract**

Considering the 21st century skills and the importance of STEM education in fulfilling these skills, it is clear that the course materials should be materials that bring students together with technology and attract their attention, apart from traditional materials. In addition, in terms of the applicability of these materials, it is very important that the materials are affordable and easily accessible. In this study two open ended resonance tube, Computer and speaker for generate sound with different frequencies, Arduino UNO, AR-054 Sound Sensor, Green LED and 220 Ω resistance were used for measure the speed of sound in air at room tempature. With the help of sound sensor, two consecutive harmonic frequency values were determined and the fundamental frequency was calculated. Using the tube features and the fundamental frequency value, the speed of sound propagation in the air at room temperature was calculated as 386.42 m/s. This value is theoretically 346 m/s. This study, in which the propagation speed of the sound is calculated with very low cost and coding studies with 12% error margin, is important in terms of hosting all STEM gains and can be easily applied in classrooms.

## 1. Introduction

The equipment which is used for measuring variables in physics experiments are usually very expensive or hard to provide. Consequently, this may cause economic or educational problems. The mentioned equipment contains a sensor to measure a specific variable. Instead of buying expensive measuring equipment, we can prepare ours at remarkably low cost. We just need a specific sensor, Arduino ™ and code.  Arduino is an electronic equipment which has inputs and outputs and many sensors that can be use with[1]. There are many experiments conducted with the aim of measuring several variables with Arduino[2-4].

In one study Arduino have used in order to measure the speed of sound. Unluckily, the path in this study is different from the usual[5]. In addition, many experiments have been conducted to measure the speed of sound. They use both open ended and one open ended resonance tubes[6-8]. The speed of sound can be calculated by changing the frequency with a stable distance or changing the distance with a stable frequency. In this studies, the resonance point is only detected aurally, not by any measurement equipment. Since detecting the highest value of resonance only aurally is very difficult, there may be mistakes in the experiment. Using measurement equipment to detect resonance value will greatly reduce the margin of error.

## 2. Experimental Setup

Within the framework of this study, sound sensor is connected to 7[th] pin and the green LED and 220 Ω resistance are connected to 8[th] pin of Arduino as shown in Figure 1.

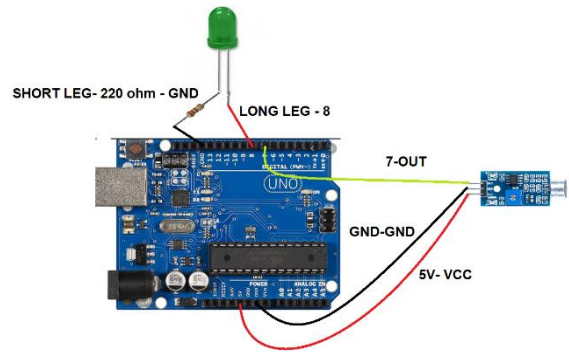

**Figure 1**. Arduino circuit prepared with sound sensor and LED

The sound sensor connected to the Arduino makes digital readings and the sensor gives "1" value when the volume is high and "0" when the volume is low. Arduino is programmed to read the data coming through the sound sensor and power it on the LED when the sensor gives the value of 1, and not power when it gives the value of 0. The two open ended resonance tube (L = 31,5 cm, r = 6,1 cm) is placed on the table horizontally. On one end of the tube, the sound sensor which is connected to the Arduino is placed and on the other end the speaker is placed as in Figure 2.

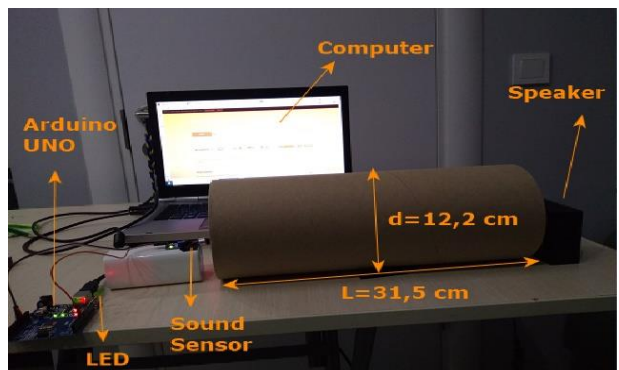

**Figure 2.** Materials and experimental setup. Arduino circuit is placed on one end of the tube with two ends open and the speaker is placed on the other end. The LED on the Arduino is positioned to be easily observed.

By using an online sound generator program on computer[9], the sound frequency is increased. At the beginning, LED does not flash 3.a. As the frequency increases, LED stars emitting light at some values as in Figure 3.b.

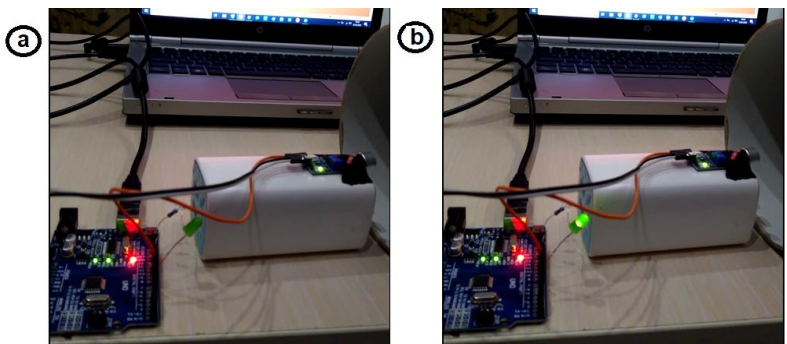

**Figure 3.** (a) Apart from harmonic frequency values, there is no resonance in the tube. Therefore, the loudness is low, the sound sensor reads 0 and no current is sent over the LED. (b) When the frequency value of the sent sound is equal to one of the harmonic frequencies, resonance occurs in the tube and the sound sensor starts to output 1. In these cases, current is sent over the LED and the LED starts scattering.

As the frequency is increased, resonance occurs in the tube at some frequency values. At these values, the sound intensity increases in the tube. And in these cases, current is passed over the LED and the LED emits light thanks to the code loaded on the Arduino. These frequency values are called harmonic frequency. The relationship between the frequency value which is measured in the two open ended tube ($f_h$), and fundamental frequency ($f_0$) is as follows.

$$f_h = (n+1)f_0 \ , \ (n=0, 1, 2....) \quad (1)$$

İn this equation, n is the sequence number of the harmonic frequency. If n value of harmonic frequency not known, It's not possible to obtain $f_0$ from equation 1. In this case, the difference between two consecutive resonance frequencies equals the fundamental frequency $f_0$.

$$f_{k+1} - f_k = (k+1+1)f_0 - (k+1)f_0 \quad (2)$$

$$f_{k+1} - f_k = f_0 k + 2f_0 - f_0 k - f_0 \quad (3)$$

$$f_{k+1} - f_k = f_0 \quad (4)$$

At the fundamental frequency, while standing waves are formed and L for the length of the tube, the wavelength λ, equals 2L. As Levine and Schwinger's study[10] that proves the tubes acoustic length is longer than the physical length, this equation becomes to

$$\lambda = 2(L + 2\alpha). \quad (5)$$

In this equation, α depends on r radius and α=0.61r. The most basic velocity formula that used for stable waves in two open ended resonance tube can be written as

$$v = \lambda f_0 = 2f_0(L + 2\alpha). \quad (6)$$

## 3. Data Analysis

At data collection process frequency value is increased and two consecutive harmonic frequencies were determined. These values are 420 Hz and 916 Hz respectively. With using these values, the fundamental frequency was calculated as 496 Hz.

If equation 3 is calculated using known values (L=0.315 m, r=0.0611 m, $f_0$=496 Hz) the speed of sound is found as 386.42 m/s at room temperature. Theoretically, speed of the sound in the air at room temperature 346 m/s. Theoretical and calculated values are extremely close to each other. The percent error is calculated as 12%. This result has indicates that Arduino can be used for measuring the speed of the sound in the air, as a very cheap, simple and effective option.

## 4. Conclusion

The velocity of the sound can be measured with many different methods. Sometimes, in some under-equipped schools, harmonic frequency values is detected only aurally. But the hearing frequency values may about ±30 Hz from the exact harmonic frequency value. Unfortunately, this reduces the precision of the study. To prevent it, Arduino and sensors can be used as cheap, simple, effective and sensitive measurement equipment. In this study, the speed of sound is measured with only 11% tolerance. This study shows that arduino is very useful for preparing lesson environments suitable for 21st century skills and performing inexpensive, high-efficiency classroom experiments.

Code:

```
void setup() {
Serial.begin(9600);
pinMode(7, INPUT);
pinMode(8,OUTPUT);
}
void loop() {
  if(digitalRead(7)==0)  // when there is sound coming to the sensor
  { digitalWrite(8, LOW);  }
  else  //  when there is no sound coming to the sensor
  { digitalWrite(8, HIGH);  }
 }
```